\documentclass[twocolumn,table,dvipsnames]{article}
\usepackage[utf8]{inputenc}
\usepackage{setspace}
\usepackage{graphicx}
\usepackage{amssymb}
\usepackage{amsmath}
\usepackage{graphicx}
\usepackage{latexsym}
\usepackage{url}
\usepackage{fancyvrb}
\usepackage{subcaption}
\usepackage{paralist}
\usepackage{listings}
\usepackage{mathtools}
\usepackage[skip=0.5pt]{caption}
\usepackage[vlined,ruled,commentsnumbered]{algorithm2e}
\usepackage{tikz,tkz-graph}
\usepackage{pifont}
\usetikzlibrary{trees,topaths,calc,shapes.geometric,arrows.meta,positioning,intersections,fit,decorations.pathreplacing}
\usepackage{graphicx}
\usepackage{setspace}
\usepackage{subcaption}
\usepackage{listings}
\usepackage{tikz}
\usetikzlibrary{matrix}
\usepackage{hyperref}
\hypersetup{
    colorlinks=true,
    linkcolor=blue,
    filecolor=magenta,      
    urlcolor=blue,
}

\colorlet{circle edge}{blue!50}
\colorlet{circle area}{blue!20}

\tikzset{filled/.style={fill=circle area, draw=circle edge, thick},
    outline/.style={draw=circle edge, thick}}

\tikzset{
    every node/.style={font=\sffamily\small},
    main node/.style={thick,circle,draw,font=\sffamily\Large}
}

\makeatletter
\def\thm@space@setup{%
  \thm@preskip=2\topsep \thm@postskip=\thm@preskip
}
\makeatother
 
\definecolor{keywords}{rgb}{0,0,0.7}

\newcommand{\cut}[1]{}

\DeclareMathOperator*{\argmax}{arg\,max}

\newcommand{\sysName}{{\tt T-REx}} %Table repair explanations
\newcommand{\holo}{{\tt HoloClean}}

\newtheorem{theorem}{Theorem}[section]

\newtheorem{example}[theorem]{Example}

\AtBeginDocument{%
  \providecommand\BibTeX{{%
    \normalfont B\kern-0.5em{\scshape i\kern-0.25em b}\kern-0.8em\TeX}}}

\makeatletter
\def\and{%
  \end{tabular}%
  \hskip 0.7em \@plus.17fil\relax
  \begin{tabular}[t]{c}}
\makeatother

\title{\sysName: Table Repair Explanations}

\author{
  Daniel Deutch \\ \small{Tel Aviv University} \\ \small{danielde@post.tau.ac.il}
  \and
  Nave Frost \\ \small{Tel Aviv University} \\ \small{navefrost@mail.tau.ac.il}
  \and
  Amir Gilad\\ \small{Tel Aviv University} \\ \small{amirgilad@mail.tau.ac.il}
  \and
  Oren Sheffer \\\small{Tel Aviv University} \\ \small{orensheffer@mail.tau.ac.il}
}
\date{}

\begin{document}
\maketitle

\begin{abstract}
Data repair is a common and crucial step in many frameworks today, as applications may use data from different sources and of different levels of credibility. Thus, this step has been the focus of many works, proposing diverse approaches. To assist users in understanding the output of such data repair algorithms, we propose \sysName, {\em a system for providing data repair explanations through Shapley values}. The system is generic and not specific to a given repair algorithm or approach: it treats the algorithm as a black box.  Given a specific table cell selected by the user, \sysName\ employs Shapley values to explain the significance of each constraint and each table cell in the repair of the cell of interest. \sysName\ then ranks the constraints and table cells according to their importance in the repair of this cell. 
This explanation allows users to understand the repair process, as well as to act based on this knowledge, to modify the most influencing constraints or the original database. 
\end{abstract}

\section{Introduction}\label{sec:intro}
Multiple previous works have proposed algorithms for data repair using Denial Constraints (DCs) \cite{DiscoveringChuIP13} or subsets thereof \cite{RekatsinasCIR17,VolkovsCSM14,ChuIP13,BohannonFGJK07}. 
These approaches employ algorithms that use the constraints to detect and change values in a database table. 
We propose {\em a system that provides explanations for data repairs by presenting the influence of each constraint and table cell.} 
An explanation for such a repair may be useful both as means of understanding the repair process and algorithm, and as a tool for debugging the quality of the constraints for the repair of this specific data.

\sysName \footnote{Please refer to the video of the system at \textit{\color{blue}\url{https://youtu.be/xPVWzHPOuAk}}} is a novel system for data repair explanations based on {\em Shapley values} \cite{shapley1953value}. 
The notion of Shapley values was originally suggested in the context of Game Theory as a measure of quantifying the contribution of each player in a cooperative game. It was later adopted by the Machine Learning (ML) community as a tool for evaluating the contribution of each feature in the model \cite{LundbergL17}. 
Given a repaired cell, \sysName\ computes and presents the Shapley values of the DCs and table cells that have influenced this repair. Our approach evaluates the contribution of the input directly rather than the contribution of hidden features which are used by a specific algorithm. {\em This allows our solution to treat the repair algorithm as a black box and only query it to compute the Shapley values of DCs and cells.} 
Explanations for the influence of DCs on the repair may assist users in correcting them and adapting them to the specific data and repair algorithm, while explanations about the influence of data cells can help in understanding the repair algorithm itself and changing specific cells to make the repair more accurate.

\begin{figure}[]
\begin{scriptsize}
 \begin{lstlisting}[mathescape=true, basicstyle=\linespread{1.5}]
$\frac{1}{6}$: (C1) $\forall t_1, t_2.~ \neg (t_1[Team] = t_2[Team] \land t_1[City] \neq t_2[City])$
$\frac{1}{6}$: (C2) $\forall t_1, t_2.~ \neg (t_1[City] = t_2[City] \land t_1[Country] \neq t_2[Country])$
$\frac{2}{3}$: (C3) $\forall t_1, t_2.~ \neg (t_1[League] = t_2[League] \land$
            $t_1[Country] \neq t_2[Country])$
0: (C4) $\forall t_1, t_2.~ \neg (t_1[Team] \neq t_1[Team] \land  t_1[Year] = t_1[Year] \land$ 
            $t_1[League] = t_2[League] \land t_1[Place] = t_2[Place])$
\end{lstlisting}
\end{scriptsize}
\caption{Denial constraints with their Shapley value}\label{fig:dcs}
\end{figure}

\begin{figure*}
    \centering
    \begin{subfigure}{0.48\textwidth}
        \includegraphics[width=\textwidth]{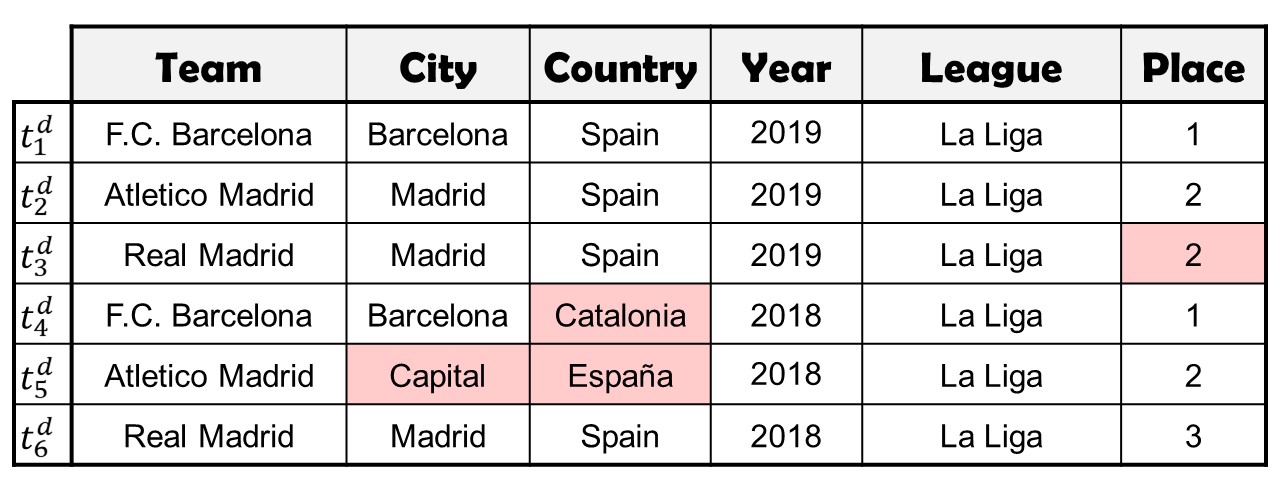}
        \caption{Dirty table (red cells are dirty)}
        \label{fig:dirtyTable}
    \end{subfigure}\hfill%
    \begin{subfigure}{0.48\textwidth}
        \includegraphics[width=\textwidth]{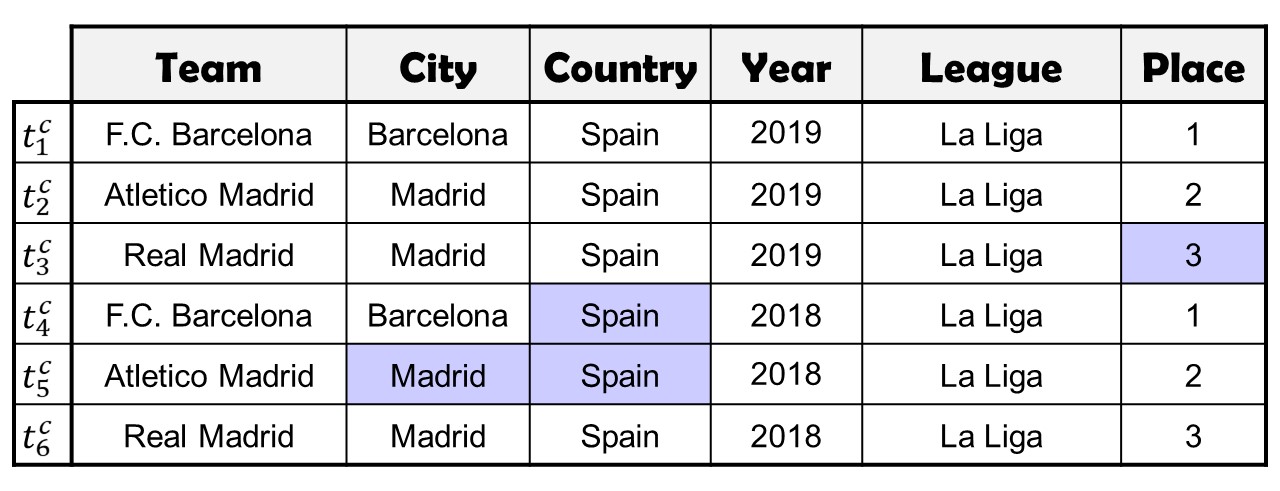}
        \caption{Clean table  (blue cells have been repaired)}
        \label{fig:cleanTable}
    \end{subfigure}
    \caption{Input dirty table and output clean table for La Liga standings}
    \label{fig:running}
\end{figure*}

\begin{algorithm}[t]
\begin{small}
    \SetKwInOut{Input}{Input}\SetKwInOut{Output}{Output}
    \LinesNumbered
    \Input{Set of constraints $\mathcal{C}$, a dirty database table $T^d$}
    \BlankLine
    \begin{enumerate}
    \item 
    If tuple $t$ has a contradiction according to $C1$ then the $City$ attribute will be modified to the most common one, i.e., $\argmax_c \mathbb{P} \left[ City = c \right]$. 
    
    \item
    If tuple $t$ has a contradiction according to $C2$ then the $Country$ attribute will be modified to the most probable one given $t \left[ City \right]$. 
    i.e., $\argmax_c \mathbb{P} \left[ Country = c \mid City = t \left[ City \right] \right]$. 
    
    \item
    If tuple $t$ has a contradiction according to $C3$ then the $Country$ attribute will be modified to the most common one, i.e., $\argmax_c \mathbb{P} \left[ Country = c \right]$. 
    
    \item
    If tuple $t$ has a contradiction according to $C4$ then the $Place$ attribute will be modified to the most probable one given $t \left[ Team \right]$, i.e., $\argmax_p \mathbb{P} \left[ Place = p \mid Team = t \left[ Team \right] \right]$. 
    \end{enumerate}
\caption{\small Simple Repair Algorithm}
\label{algo:repair}
\end{small}
\end{algorithm} \DecMargin{1em}
\vspace{-1mm}

\begin{example}
Consider the table in Figure \ref{fig:dirtyTable} and the DCs in Figure \ref{fig:dcs} with the Shapley values of each DC on its left.
C1 says that two tuples that share a team value must be in the same city, C2 says that if a pair of tuples share a city, they must have the same country, C3 says that two tuples that have the same league must have the same country, and C4 says that it is impossible for two different teams of the same league to finish in the same place in the same year. 
Consider the cell $Country$ in the fifth row, denoted by $t_5[Country]$.
For simplicity, assume that we have Algorithm \ref{algo:repair} as a n\"aive repair algorithm\footnote{In practice, the repair algorithm may be more sophisticated; our solution is agnostic to the complexity of the repair algorithm.}. 
\sysName\ computes the contribution of each DC and ranks them accordingly, where C3 is the most influential DC. It contributed the most as the $League$ value ``La Liga'' appears in 3 other tuples coupled with the value ``Spain'' in the attribute $Country$. C1 and C2 each contributed equally as C1 caused the change of ``Capital'' to ``Madrid'' first and then C2 caused the change of the value in the $Country$ cell. C4 is not involved in the repair so its contribution is $0$.

Next, we measure the influence of different data cells on this repair.
Given Algorithm \ref{algo:repair}, observe that the value of $t_1[Place]$ has no influence on the modification of $t_5[Country]$ -- as $t_1$ has no contradictions with $t_5$, and the attribute $Place$ does not affect $Country$ in Algorithm \ref{algo:repair}. However, how can we determine if $t_5[League]$ was more or less influential on the repair compared to $t_6[City]$? 
Intuitively, $t_5[League]$ is more influential than $t_6[City]$. 
This is because if $t_5[League]$ had a different value, then tuple $t_5$ would not have any contradictions according to $C3$. While if $t_6[City]$ had a different value, then according to $C1$ there would have been a contradiction between $t_3$ and $t_6$ (as both tuples would have $Team$ value of ``Real Madrid", and an inconsistent $City$) which would have been resolved by Algorithm \ref{algo:repair}.
As a result \sysName\ will assign higher contribution to $t_5[League]$ compared to $t_6[City]$.
\end{example}

\sysName\ takes as input the algorithm itself and its input which is a set of DCs and a dirty database table. Another input to the system is a specific table cell of interest whose repair requires explaining. 
The system then ranks the influencing DCs and table cells based on their Shapley value for this cell of interest. 
Generally, computing the Shapley value is exponential time in the number of DCs/table cells, and thus \sysName\ employs different algorithms to compute the Shapley value for DCs and for table cells. 
With DCs, the n\"aive approach is feasible as the number of DCs is usually small. 
Conversely, the number of cells in a table can be very large, so \sysName\ uses a sampling algorithm based on \cite{StrumbeljK14}. 
To compute the Shapley values, the system repeatedly changes the input of the repair algorithm and  queries it, so it does not rely on the components or approach of a specific algorithm. 
\section{Technical Details}\label{sec:tech}
We give a short overview of the approach underlying \sysName. 

\subsection{Database Repair}
$T$ will denote a database table with schema $(A_1, \ldots, A_m)$ where $A_i$ is the $i$th attribute of $T$. For a tuple $t\in T$, the notation $t[A_i] = v$ means that $t$ has the value $v$ in attribute $A_i$. We denote by $T^d$ and $T^c$ the database table prior to the repair and after it respectively. Extending this, $t^d[A]$ and $t^c[A]$ will also be used to denote a dirty and clean cell, respectively.

\begin{example}
Consider the dirty and clean tables shown in Figures \ref{fig:dirtyTable}, \ref{fig:cleanTable}, referred to as $T^c$ and $T^d$. If we consider $t_5$ in both tables, then the attribute $t_5^d[Country]$ in $T^d$ is changed in $T^c$ from the value ``Espa\~na'' to ``Spain''.
\end{example}

We denote the repair algorithm by $Alg$ and its input by (1) $\mathcal{C}$, a set of DCs and (2) $T^d$, a dirty table. Also, denote $Alg(\mathcal{C}, T^d) = T^c$ as the output table of $Alg$. 
For our purposes, we will refer to $Alg$ as a binary function as follows. 
Given a table cell $t^d[A]\in T^d$, the repair algorithm is a function $Alg|_{t^d[A]}:(\mathcal{C}, T^d) \to \{0,1\}$, where $1$ signals that the value in $t^d[A]$ is repaired to the value in $t^c[A]$, and $0$ otherwise. 

\begin{example}
Consider the cell $t_5[City]$ in Figures \ref{fig:dirtyTable} and \ref{fig:cleanTable}. Without C1 it would not have changed from ``Capital'' to ``Madrid'', therefore: $Alg|_{t_5[C]}(\{C1, C2, C3\}, T^d) = 1$ while $Alg|_{t_5[C]}(\{C2, C3\}, T^d) = 0$.
\end{example}

\subsection{Shapley Value}
In Cooperative Game Theory, Shapley value \cite{shapley1953value} is a way to distribute the worth of all players, assuming they cooperate. 
Let $N$ be a finite set of players and $v : 2^N \to \mathbb{R}$, $v(\emptyset) = 0$ be a function (called a characteristic function). $v$ maps sets of players to the joint worth they generate according to the game. The Shapley value of a player $a$ is then defined as:
\vspace{-1.5mm}
\begin{scriptsize}
\begin{equation*}
\begin{split}
    Shap(N,v,a) = \sum_{S\subseteq N\setminus \{a\}} \frac{|S|!(|N|-|S|-1)!}{|N|!}\cdot (v(S\cup \{a\}) - v(S))
\end{split}
\end{equation*}
\end{scriptsize}
In our scenario, the model is a black box so the Shapley values are computed on the input itself, i.e., the constraints and the table. 
For constraints, we adapt the definition so that it reflects the contribution of a specific constraint to the repair of a cell, as follows.
\vspace{-1.5mm}
\begin{scriptsize}
\begin{equation*}
\begin{split}
Shap(\mathcal{C},Alg|_{t^d[A]},C) =  \sum_{\makebox[0pt]{$S\subseteq \mathcal{C}\setminus \{C\}$}} \frac{|S|!(|\mathcal{C}|-|S|-1)!}{|\mathcal{C}|!}\cdot (Alg|_{t^d[A]}(S \cup \{C\},T^d) -\\ Alg|_{t^d[A]}(S,T^d))
\end{split}
\end{equation*}
\end{scriptsize}
Where $t^d[A]$ is a specific cell of interest and $C$ is a constraint whose contribution we want to determine. The ``set of players'' is the set of DCs while the table $T^d$ remains constant.

\begin{example}\label{ex:ic}
Recall the tables in Figure \ref{fig:running} with the DCs in Figure \ref{fig:dcs} (Shapley values are on the left) and Algorithm \ref{algo:repair}. 
We now compute the contribution of each DC to the repair of the cell $t_5[Country]$, denoted $t_5[C]$. 
Algorithm \ref{algo:repair} will repair $t_5[C]$ only if we have the DCs $\{C1,C2\}$, or $\{C3\}$.  
According to the definition, we can compute the contribution of $C_1$ as follows: there are 8 subset of $\{C2,C3,C4\}$, and only for $S = \{C2\}$ and $S = \{C2, C4\}$ we have $Alg|_{t_5[C]}(S\cup\{C1\},T^d) = 1$ and $Alg|_{t_5[C]}(S,T^d) = 0$, so $Shapley(\mathcal{C}, T^d, C1) = \frac{2}{12}$. The same computation applies to $C2$. 
For $C3$ we have 6 out of 8 subsets $S$ of $\{C1,C2,C4\}$ that result in $Alg|_{t_5[C]}(S\cup\{C3\},T^d) = 1$ and $Alg|_{t_5[C]}(S,T^d) = 0$, including $S = \emptyset$. Thus, $Shapley(\mathcal{C}, T^d, C3) = \frac{2}{3}$. 
As for $C4$, its presence or absence does not change the value of $t_5[C]$, so $Shapley(\mathcal{C}, T^d, C4) = 0$. 

Let us explain the intuition for the value of $C3$ being double that of the pair $\{C1, C2\}$. Ignore for now $C4$ since its contribution is $0$. There are $5$ subsets of the DCs $\{C1, C2, C3\}$ for which we repair $t_5[C]$. These are $\{C3\}$,  $\{C1, C2\}$,  $\{C1, C3\}$,  $\{C2, C3\}$, and $\{C1, C2, C3\}$.
Four of these sets contain $C3$ while only two contain the pair $\{C1,C2\}$ (for the subsets where one of these is present without its partner, the repair is due to $C3$), 
thus, the contribution of $C1$ and $C2$, as a pair, is half that of $C3$.
\end{example}

Similarly, we adjust the definition for the Shapley value of a cell. 
Given a repair of cell $t^d[A]$ we define the formula for calculating the Shapley value of a cell $t_i[B]$, or intuitively, its contribution to the repair of $t^d[A]$.
\vspace{-1.5mm}
\begin{scriptsize}
\begin{equation*}
\begin{split}
Shap(D,Alg|_{t^d[A]},t_{i}[B]) = \sum_{\makebox[0pt]{$S\subseteq T^d\setminus \{t_{i}[B]\}$}} \frac{|S|!(|T^d|-|S|-1)!}{|T^d|!}\cdot \\ 
{} {} (Alg|_{t^d[A]}(\mathcal{C},S\cup \{t_{i}[B]\}) - Alg|_{t^d[A]}(\mathcal{C},S))
\end{split}
\end{equation*}
\end{scriptsize} 
Where $S\subseteq T^d$ means $\forall t_j[C] \in T^d\setminus S.~ t_j[C] = null$. 
Here, the ``set of players'' here is the set of cells in the table $T^d$ while the set of constraints remains constant.

\begin{example}
Reconsider our example with the DCs from Figure \ref{fig:dcs}, Algorithm \ref{algo:repair}, and the tables in Figure \ref{fig:running}. Consider the cell $t_5[Country]$ whose value is changed from ``Espa\~na'' to ``Spain''. 
Among all the cells, $t_5[League]$ has the highest Shapley value, next we will explain why. Notice that based on C3 the inclusion of $t_5[League]$ to any coalition that contains at least one of the pairs $\{t_i[Country], t_i[League]\}$ for any $i\in \{1, 2, 3, 6\}$ would result in the repair of $t_5[Country]$ to ``Spain". Observe that there are $175 \cdot 2^{27}$ such coalitions (since out of the relevant $8$ cells there are $2^8-3^4=175$ options to choose a coalition such that at least one pair exists, and excluding those cells and $t_5[League]$ there are $27$ remaining cells that can be either included or excluded from the coalition).
Next, we will estimate the number of coalitions that are required for the fix based on C1 and C2. According to these DCs, a coalition that contains $\{t_3[Team], t_3[City], t_3[Country], t_5[Team]\}$ is required. There are $2^{32}$ such coalitions.
Since $175 \cdot 2^{27}$ is more than five times larger than $2^{32}$ we conclude that $t_5[League]$ has the highest influence on the repair of $t_5[Country]$ from ``Espa\~na'' to ``Spain''.
For simplicity we overlooked the coalitions sizes, though they too play a role in the evaluation of Shapley values. 
\end{example}

\subsection{Computing Shapley Values}
Shapley values can be computed from the definition, but the computation time may be exponential. 
For constraints, we can use the formula directly as their number is typically small. 
However, the number of table cells can be huge. 
Therefore, we use a novel algorithm based on probabilistic sampling \cite{StrumbeljK14} to approximate the contribution of a table cell.

\begin{example}
Reconsider the table in Figure \ref{fig:dirtyTable}. Suppose we are interested in the effect of the cell $t_5[City]$ on the repair of the cell $t_5[Country]$. We initialize a variable $\varphi = 0$. 
We vectorize the table 
to get the vector $x_T = (t_1[Team], t_1[City], \ldots, t_2[Team],\\ \ldots, t_6[Place])$. 
To sample a cell coalition, we take a random permutation of $x_T$-- the coalition is the set of all of the cells that precede $t_5[City]$.
Values of cells that are not part of the coalition will be replaced with a sample value from their column distribution. 
Once the cell coalition was formed we generate two instances of vectorized tables: one with the original value of $t_5[City]$, and the second where the $t_5[City]$ value is replaced with random value. 
We then compute the difference in the result of $Alg|_{t_5[Country]}$ for these two instances and add it to $\varphi$. We repeat this $m$ times and output $\frac{\varphi}{m}$.
\end{example}

\begin{figure*}[!htb]
    \centering
    \begin{subfigure}{.3\linewidth}
    \includegraphics[width=\linewidth]{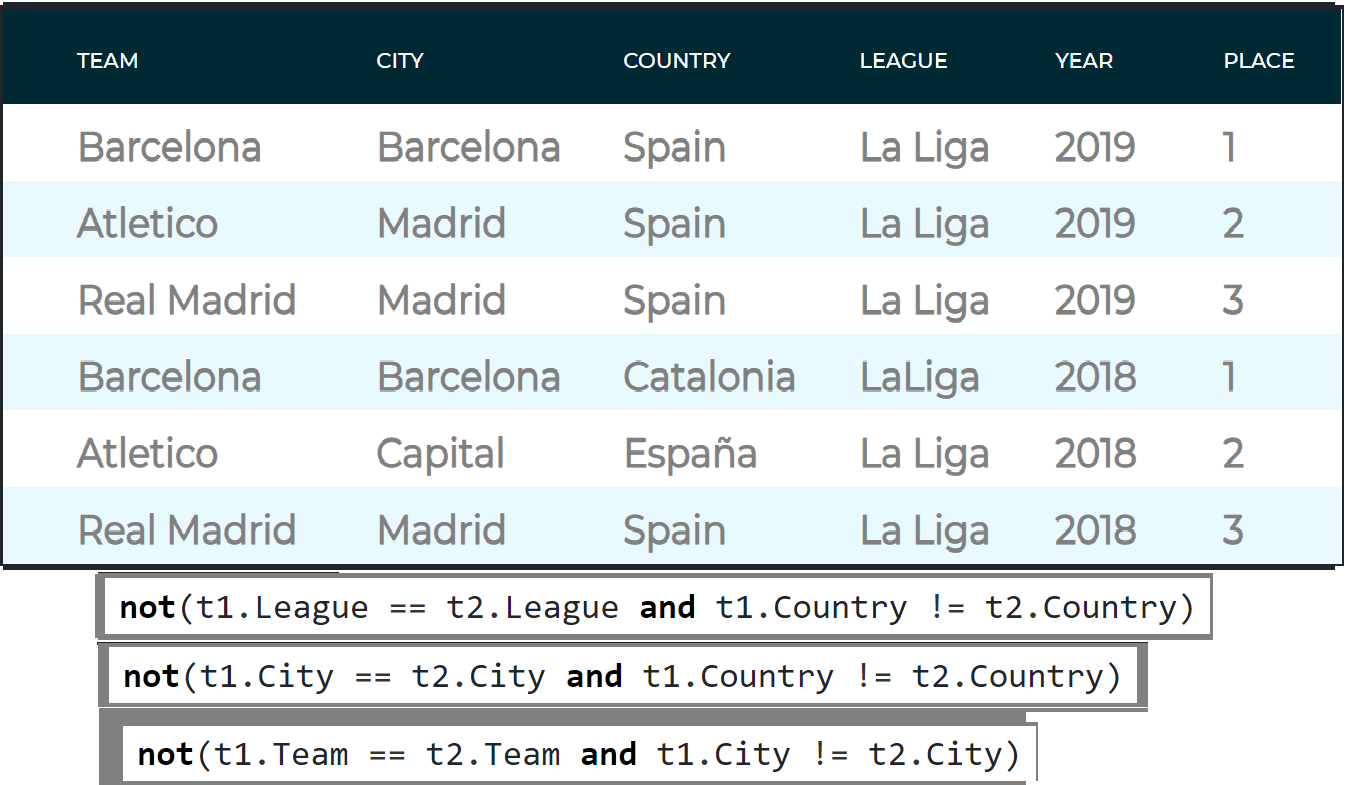}
    \caption{Input Screen}\label{fig:sys_input}
    \end{subfigure}
    \begin{subfigure}{.3\linewidth}
    \includegraphics[width=\linewidth]{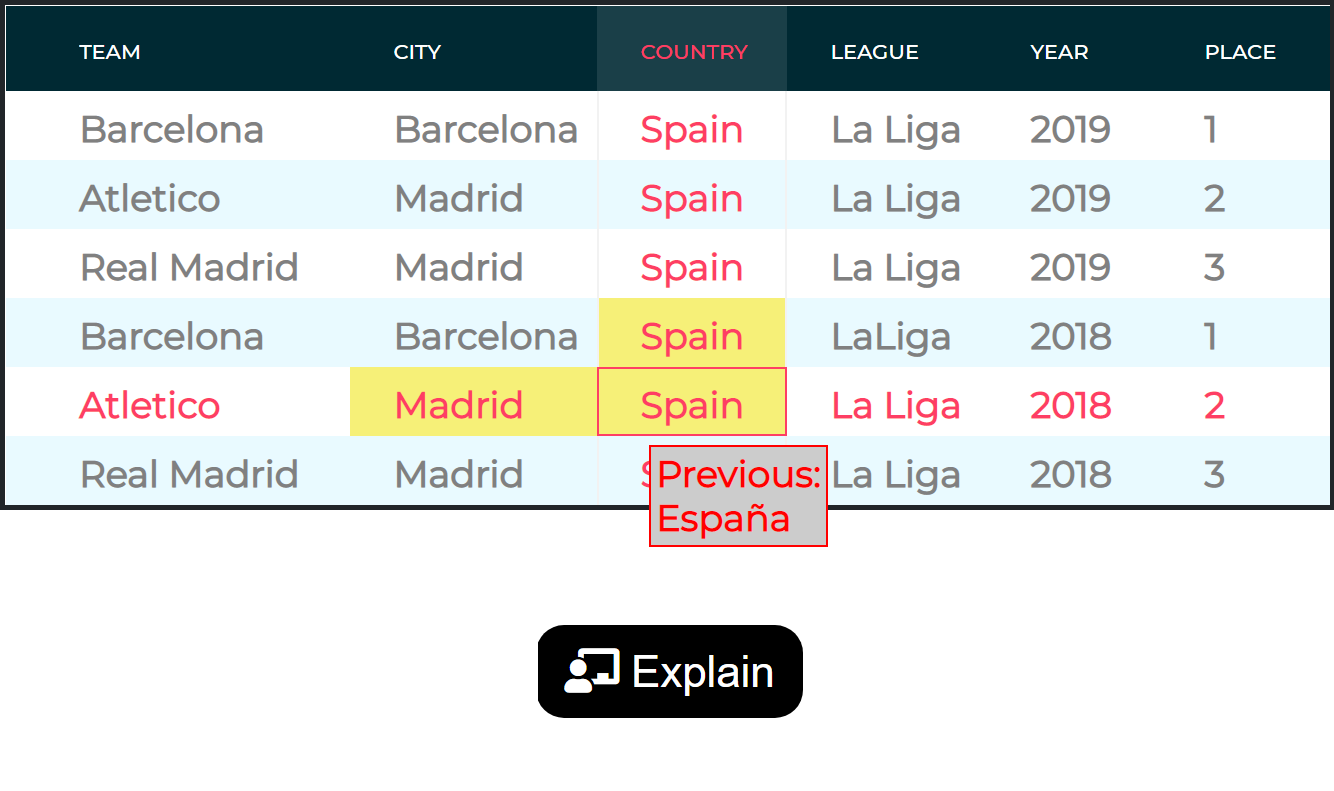}
    \caption{Repair Screen}\label{fig:sys_repair}
    \end{subfigure}
    \begin{subfigure}{.3\linewidth}
    \includegraphics[width=\linewidth]{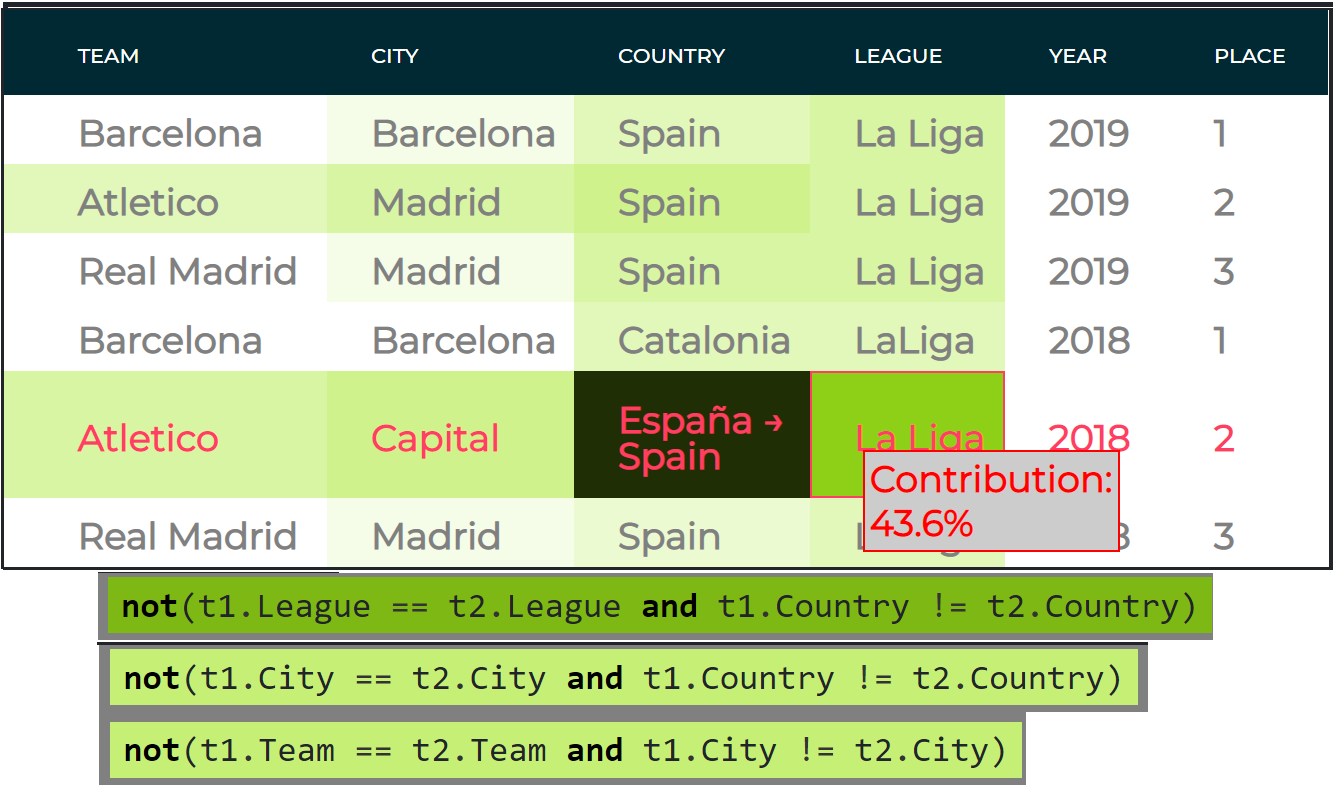}
    \caption{Explanation Screen}\label{fig:sys_exp}
    \end{subfigure}
    \caption{\sysName\ User Interface}\label{fig:sys_ui}
\end{figure*}

\begin{figure}
    \centering
    \includegraphics[width=0.4\textwidth]{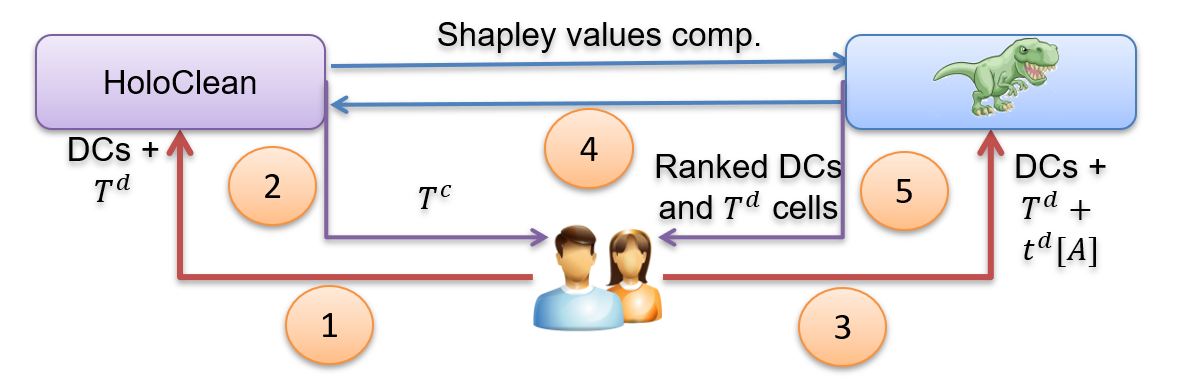}
    \caption{\sysName\ Architecture}\label{fig:framework}
\end{figure}

\section{System Overview}\label{sec:system}
\sysName\ is implemented in Python 3.6 and an underlying database engine in PostgreSQL 10.6. Its web-based GUI was built using JavaScript, CSS and HTML. 
The three screens of the system are shown in Figure \ref{fig:sys_ui} and the general architecture of \sysName\ is shown in Figure \ref{fig:framework}. 
Users first input a database table and a set of DCs to the \holo\ system (Figure \ref{fig:sys_input} and the arrow 1 in Figure \ref{fig:framework}). \holo\ \cite{RekatsinasCIR17} is a holistic data repair system, that supports DCs, among other forms of constraints, and repairs the input table based on a probabilistic model involving machine learning techniques. 
After clicking the ``Repair'' button, users are presented with the repaired table, where repaired cells are highlighted (Figure \ref{fig:sys_repair}). Furthermore, when hovering over a repaired cell, the system shows its value before the repair. 
Now, \sysName\ allows users to choose any cell, $t^d[A]$, from the original table, $T^d$, whose value was changed, and mark it as a cell of interest and click the ``Explain'' button. 
The system then computes the Shapley values w.r.t. the chosen options by querying \holo\ as part of the computation. 
Once done, \sysName\ displays the DCs and table cells ranked from highest to lowest in terms of their Shapley value w.r.t. $t^d[A]$, where influencing DCs and cells are highlighted green and the darker the color, the more influencing the DC/cell is (Figure \ref{fig:sys_exp}). Again, when hovering over the DCs/cells users can also see their Shapley values. 
The user can continue the process by changing the DCs or values in $T^d$, and inputting it again to \holo\ to infer another repair, thus improving the repair iteratively.

\section{Demo Scenario}\label{sec:scenario}
Our demonstration will show that explaining repairs through Shapley values assists in understanding the repair process and debugging it. 
We will use a soccer database, scraped from Wikipedia, similarly to Figure \ref{fig:dirtyTable}, and errors will be manually added into the table. 
We will start with an initial set of DCs. 
To get the repair, we will employ \holo\ that will output a clean table. Then, we will indicate a repaired cell of interest and show the most influential table cells and DCs involved in this repair, ranked according to their Shapley value. 
We will show how removing or changing the highest ranked DCs improves the repair of the specified table cell. 
We will use a similar scenario for table cells, where the DCs will be appropriate but some of the cells will cause a specific cell to be repaired in the wrong manner. After showing the obtained repair, we will invoke \sysName\ to rank the influencing table cells. 
We will then allow users to change values in the initial table and the DCs and choose different cells of interest to them. 
Users could then use \sysName\ to compute the Shapley value of the table cells and DCs that influenced the repair of their chosen cell and explore the system. 

\paragraph{Acknowledgements}
\small{
This research has been funded by the European
Research Council (ERC) under the European Union’s Horizon 2020
research and innovation programme (Grant agreement No. 804302),
the Israeli Science Foundation (ISF) Grant No. 978/17, and the Google
Ph.D. Fellowship. The contributions of Nave Frost and Amir Gilad are part of their respective Ph.D. thesis research conducted at Tel Aviv University.}

\clearpage
\bibliographystyle{abbrv}
\bibliography{bibtex.bib}

\end{document}